\documentclass[twocolumn,showpacs,preprintnumbers,amsmath,amssymb]{revtex4}

\usepackage{graphicx}
\usepackage{dcolumn}
\usepackage{bm}

\begin{document}

\title{Mixing and condensation in a wet granular medium}

\author{Dimitrios Geromichalos}
    \email{Dimitrios.Geromichalos@physik.uni-ulm.de}

\author{Mika M. Kohonen}

\author{Frieder Mugele}

\author{Stephan Herminghaus}

\affiliation{%
    Applied Physics Lab., Ulm University, Ulm, D-89069,
    Germany
}

\date{\today}

\begin{abstract}
We have studied the effect of small amounts of added liquid on the
dynamic behavior of a granular system consisting of a mixture of
glass beads of two different sizes. Segregation of the large beads
to the top of the sample is found to depend in a nontrivial way on
the liquid content. A transition to viscoplastic behavior occurs
at a critical liquid content, which depends upon the bead size. We
show that this transition can be interpreted as a condensation due
to the hysteretic liquid bridge forces connecting the beads, and
provide the corresponding phase diagram.
\end{abstract}

\pacs{05.65.+b,45.70.-n,45.70.Mg}

\maketitle

As is generally known from everyday experience, the mechanical
properties of a granular material change dramatically if some
liquid is added. The main reason is the internal cohesion due to
capillary forces arising from liquid bridges between the grains
\cite{mik98, israelachvili, horn97, hal98, boc98, fra99}. While
recent years have seen considerable progress in understanding the
dynamics of dry granular materials \cite{ristow, jae96, kad99},
the physical mechanisms underlying the properties of wet systems
remain largely obscure. This is in part due to the inherent
complexity of these highly nonlinear systems, but also to the fact
that only few experiments have yet been conducted on this topic.

Some experiments have been recently done on the influence of
liquid on the mixing behavior of systems consisting of large and
small spheres \cite{sam00, sam01}. It was observed that as liquid
was added, the tendency to segregate, which is well known for dry
granulates \cite{ott00, herm02}, was diminished. At a certain
wetness, complete mixing occurred. In these experiments the
morphology of heaps formed by pouring mixtures of particles with
two different sizes through an orifice was investigated. In order
to have a closer look at the observed effects, we have carried out
experiments in which the agitation of the material is provided in
a more homogeneous way, thus ensuring a more or less well defined
granular temperature. Furthermore, the propensity of the system to
segregate into large and small beads can be determined
quantitatively in our setup. We observe a strong dependence of
mixing on the liquid content, and the transition to a viscoplastic
regime due to condensation of the grains into clusters, which was
reported in ref.\cite{sam01}, was clearly observed. However, the
overall behavior was found to be more complicated than previously
reported, and we are able to explain all of the observed features
within a rather simple theoretical model.

A typical sample consisted of $150 ml$ of glass beads with $R =
2.5 mm$ radius, sealed in a cylindrical jar (diameter: $80$ mm,
cf. fig. 1) together with the same volume of smaller glass beads
with radius $r$, which was varied from $50$ to $500 \mu m$, and a
well defined amount of liquid. In order to prevent
crystallization, the small beads were chosen to be slightly
polydisperse: the spread in bead size ranged from 10 to 20 $\%$.
The jars were mounted on a shaker, with the cylinder axis upright.
The shaking motion consisted of small horizontal circles with a
diameter of $5$ mm and a frequency of $20$ revolutions per second,
which corresponds to a velocity of the container walls of 0.38
m/s. When the segregation pattern did not change any more
(typically after one half hour), photographs of the jars were
taken and analyzed. Two examples are shown in Figure 1.

\begin{figure}[h]
 \includegraphics[width = 8.5cm]{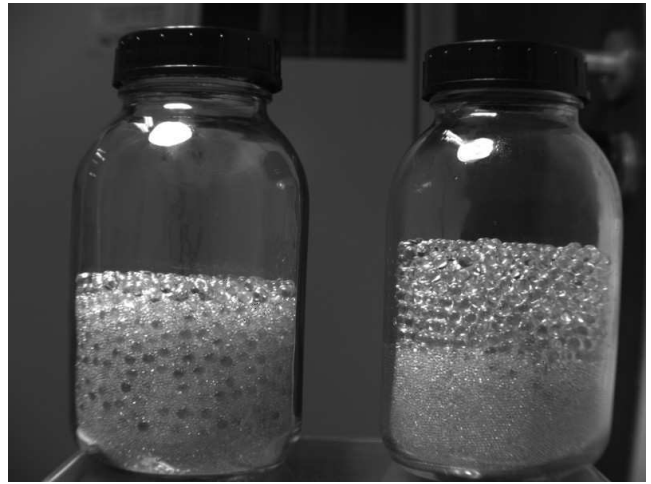}% Here is how to import EPS art
 \caption{Horizontally shaken beads (with radii $R = 2.5 mm$ and  $r = 0.5 mm$) without water (left)
 and after adding 75 $ \mu $l (right).}
\end{figure}

We defined the mixture zone as the zone in which both the smaller
and the bigger particles were visibly present. This zone was
observed to be rather homogeneous concerning the mixing ratio.
Occasional analysis of horizontal slices of the samples revealed a
small radial variation of the mixing ratio, which is not captured
by the photographs, but the results given below are qualitatively
valid for the whole volume. For the results presented here we used
water as the liquid, but similar results were obtained using
ethanol. We can therefore exclude that leaching of ions from the
glass, and other side effects due to the specific properties of
water, play a significant role \cite{oli02}. The quantity we will
discuss is the height of the mixture zone, divided by the total
height of the sample, and is denoted by $h$. For perfect
segregation, $h \rightarrow 0$, while $h = 1$ for perfect mixing.
The amount of liquid added to the samples is denoted by the
dimensionless quantity $W$, which is defined as the volume of the
liquid divided by the total volume of the small beads (including
the space between the beads).

In contrast to what one might anticipate on the basis of earlier
work \cite{sam00, sam01}, the propensity of mixing of the
particles did not improve monotonically with the amount of liquid
added. Instead, we found in most cases that the height of the
mixture zone, $h$, was first {\it reduced} upon addition of water.
Only after a certain amount of liquid had been added did the
mixing improve. In Fig. 2, $h$ is plotted as a function of $W$ for
two different $r$. The size of the large beads was kept constant.
Clearly, there is a rather well defined liquid content above which
the height of the mixing zone {\it increases} when more liquid is
added.

The boundary between these two regimes (i.e., the minima of the
curves as in fig. 2) is indicated by the open squares in Figure 3,
which represents a phase diagram of the system in the plane
spanned by the radius of the small beads, $r$, and the liquid
content, $W$. The dotted line serves as a guide to the eye. The
full circles correspond to the occurrence of complete mixing
(i.e., $h = 1$). We can thus clearly distinguish three regimes,
which we denote as {\it gaseous}, {\it intermediate}, and {\it
viscoplastic} \cite{teg99, teg02}, respectively. In what follows,
we will discuss these regimes in some detail.

\begin{figure}[h]
 \includegraphics[width = 8.5cm]{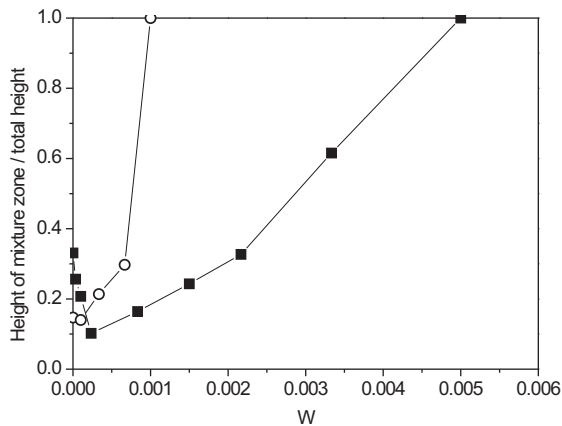}% Here is how to import EPS art
 \caption{Normalized height of the mixture zone of shaken glass beads
 as a function of the amount of water added. The radius of the big beads was $R = 5mm$,
 the diameter of the small beads was $r = 240 \mu m$ (closed symbols) and $r = 140 \mu m$
 (open symbols), respectively.}
\end{figure}

\begin{figure}[h]
 \includegraphics[width = 8.5cm]{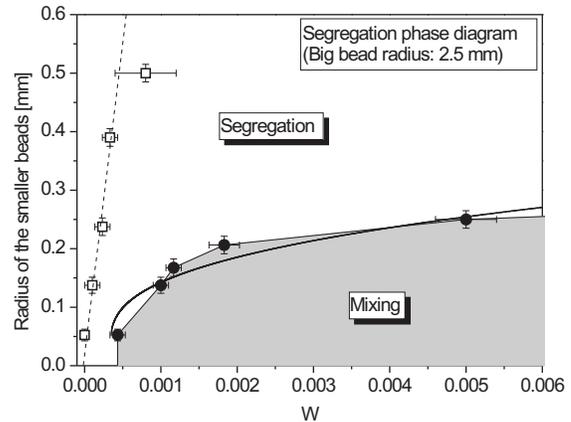}% Here is how to import EPS art
 \caption{Segregation phase diagram in dependence of the liquid added. In the grey area there
 occurs no segregation, while in the white area segregation was observed. The dashed steep line
 separates the area in which the segregation increases by adding water (left) from the area
 in which the segregation decreases or stays the same (right).}
\end{figure}

Let us start with the transition to the {\it viscoplastic regime},
in which perfect mixing is observed. This transition, which occurs
as $h \rightarrow 1$, appears quite abrupt
 when inspecting the experiment by eye.
While the general motion of the whole sample was a kind of
circular convection around the axis of the jar in the gaseous and
intermediate regimes, the motion in the case of perfect mixing was
reminiscent of kneading raisin dough. We identify this transition
with the one reported recently by other authors
\cite{sam00,sam01}. We assume that this transition into the
viscoplastic regime occurs when the granular temperature of the
system, i.e., the average kinetic energy per bead, is not anymore
 sufficient to break the liquid bridges between the beads, and is thus
 a kind of condensation within the granular gas, mediated by the liquid bridges.

In order to support this view, we have to discuss the involved
forces in some detail. The force exerted upon two adjacent beads
by the liquid bridge can be written as $F(d,w) = 2 \pi r \gamma
f(w) g(d)$, where $\gamma$ is the surface tension of the liquid,
$r$ is the radius of the spheres, $w$ is the liquid volume of the
bridge, and $f$ is some function depending on the roughness of the
sphere surfaces \cite{hal98, wil00}. $g(d)$ describes the
dependence of the force on the mutual distance, $d$, of two
ideally spherical beads. The energy required to destroy the bridge
is given by
\begin{equation}
E_{B}(w) = 2 \pi r \gamma f(w) \int_0^{d_p}g(d)dd
\end{equation}
where $d_p$ is the distance at which the bridge pinches off. The
integral can be well approximated as $w^{1/3}/3$ \cite{wil00}.

For perfect spheres and complete wetting, $f(w) = 1 \ \forall w$,
but for a rough surface, it tends to zero as $w \rightarrow 0$.
The shape of $f(w)$ has been discussed in detail before
\cite{hal98}. Based on these considerations, it can be well
approximated by
\begin{equation}
f(w)=\left[\frac{w}{w+w_0}\right]^{\mu}
\end{equation}
with $w_0 := r \delta ^2$ characterizing the roughness amplitude,
$\delta$. For the exponent, we have $\mu = (2-\chi)/(2+\chi)$,
where $\chi$ is the roughness exponent of the bead surface. For
(typical) scratch and dig roughness, $\chi$ is close to zero, and
$\mu \approx 1$. This corresponds to the regime found by Hornbaker
et al. \cite{horn97}. In the opposite case, when $\chi \rightarrow
1$, we have $\mu \rightarrow 1/3$. We see that $\mu$ is confined
to a narrow interval, and is very probably close to one in our
system.

We now must calculate the average bridge volume, $w$, from the
dimensionless liquid volume, $W$. At equilibrium, assuming that
all of the liquid is in the bridges, $w = \frac{8 \pi r^3 W}{3 c
\rho_{p} }$, where $c$ is the coordination number of the network
of bridges ($c \approx 6$) and $\rho_p$ is the packing density of
the spheres ($\rho_p \approx 0.65$). Under dynamic conditions the
typical bridge volume would be expected to be somewhat smaller
than the equilibrium value. Our measurements \footnote{M. M.
Kohonen, D. Geromichalos, and S. Herminghaus; manuscript in
preparation} suggest that $w = \alpha r^3 W$, where $\alpha
\approx 0.25$. If we now equate the bridge energy, $E_B$, to the
average kinetic energy of the small beads, $E_{K} := \frac{m}{2}
\left< v^2 \right>$, we obtain the implicit equation
\begin{equation}
r^2 - l^{1/\mu} \alpha^{1/3\mu} W^{1/3\mu} r^{(2-1/\mu)} +
\frac{\delta^2}{\alpha W} = 0
\end{equation}
where we have introduced the characteristic length scale $l :=
\frac{\gamma}{\rho \left< v^2 \right> }$ ($\rho$ is the mass
density of the bead material. In our case, $\rho = 2500 \rm
kg/m^3$). Eq. (3) can be easily solved analytically for $\mu = 1$
and $\mu = \frac{1}{3}$. For the most relevant case of $\mu = 1$,
we obtain
\begin{equation}
r = \frac{l}{2} \alpha^{1/3} W^{1/3} \left( 1 + \sqrt{1 - \left(
\frac{W_c}{W} \right)^{5/3}} \right)
\end{equation}
which intersects the W-axis at the critical moisture content $W_c
= \frac{1}{\alpha} (2\delta/l)^{6/5}$, which only depends on the
roughness amplitude, $\delta$, but not on $r$. This function
represents the phase boundary at which condensation sets in, and
is indicated as the full curve in Figure 3. From this best-fit
curve we obtain $\delta \sim 500 nm$, which is similar to the
peak-to-peak roughness obtained from the inspection of the beads
by atomic force microscopy. From the fitted value of the intrinsic
length scale $l$, and from the known density of the glass and the
surface tension of the water, we get the average granular
velocity, $v := \sqrt{\left< v^2 \right>} = 0.11 m/s$. This
compares favorably with the velocity of the shaking motion, which
is expected to be of the same order of magnitude.

Let us now turn to the {\it gaseous regime}. If there is no liquid
in the sample at all, segregation is observed as usual, which has
become known as the 'brazil nut effect' (BNE) \cite{wil76, ros87,
kni93, mob01}. As it has been shown recently \cite{herm02}, this
may be understood in a continuum picture, as follows. A granular
medium is strongly dissipative due to the inelastic character of
the impact kinetics of the grains, which is characterized by the
restitution coefficient, $\varepsilon$. The latter is defined as
the ratio of the particle velocities after and before impact,
$\varepsilon = v^{\prime}_{rel}/v_{rel}$. Thereby the granular
temperature bath, which typically corresponds to thermal
temperatures in the Terakelvin range, is internally coupled to the
room-temperature bath provided by the atoms constituting the
grains. This inherent state of non-equilibrium is the basic reason
for the peculiar features of granular matter dynamics in general.

Now consider a spherical region with radius $R$ within the
granular gas. When a grain, or small bead, enters this region, it
will have many impacts, and thereby lose a lot of energy, before
it can leave this region again. When the latter is replaced by a
solid sphere (i.e., a large bead of radius $R$), the grain will
experience only one such impact. Consequently, the granular gas is
hotter close to an intruder than far away in the bulk, because the
intruder is a less effective heat sink. The granular gas is
therefore also less dense around the intruder, and the latter
rises due to the buoyancy of the 'bubble' of reduced density it
carries around itself \cite{herm02}.

The height of the mixing zone is then a direct consequence of the
balance between the hydrostatic pressure exerted by the
large-beads fraction upon its lower interface, and the buoyancy
pressure exerted upon the same interface from below by the mixed
phase. This picture is supported by the observation that the large
beads acquire a rather uniform spacing within the mixed phase,
which might then correspond to (twice) the thickness of the
reduced density layer around each large bead \cite{herm02}.

The BNE increases as the energy loss upon impact is increased.
This, however, is just what is achieved when liquid is added. The
interparticle force exerted by the liquid has a hysteretic
character \cite{wil00, cra94, lothar02}, \footnote{M. Schulz, B.
M. Schulz, S. Herminghaus, {\it Shear-induced solid-fluid
transition in a wet granular medium}, submitted to Phys. Rev. E}
and is thus intrinsically dissipative. We thus expect the
restitution coefficient to be reduced, and the BNE to become more
prominent, when liquid is added. More precisely, the buoyancy of
the reduced density bubbles around the large beads increases, and
a shallower mixing zone is needed to balance the hydrostatic
pressure of the large-beads column above it. Thus the height of
the mixing zone is expected to decrease when liquid is added,
which is what we observe. We thus assume that throughout this
regime, which is defined by a monotonically decreasing height of
the mixing zone, the physics of two-particle impact is affected by
the liquid merely in the sense of a decrease in the restitution
coefficient, $\varepsilon$.

Although a quantitative prediction of the BNE is yet difficult, we
can give a rough estimate of the expected size of the effect. The
inelastic character of the impact of the dry particles reflects
itself in the BNE (and thus the tendency of demixing) of the dry
material. Consequently, we expect the contribution due to the
liquid to be of the same order of magnitude when the energy stored
in the liquid bridges is similar to the energy lost upon impact,
i.e., when $E_{B} \approx (1-\varepsilon^2) E_{K}$. Since $l
\propto E_{K}^{-1}$, the corresponding moisture content,
$W_{BNE}$, is related to $W_c$ via $W_{BNE} \approx
(1-\varepsilon^2)^{6/5} W_c$. Exploiting that $W_{BNE}$ roughly
corresponds to the open symbols in Fig. 3, we can estimate the
restitution coefficient of our system. We obtain $\varepsilon =
0.95 \pm 0.03$, which compares favorably with values known from
the literature \cite{ristow}.

{\it Intermediate regime:} As more liquid is added, the width of
the hysteresis loop becomes larger, such that the kinetic energy
of the impact may sometimes not be sufficient to break a bridge.
It thus appears likely that small clusters will form which have a
certain lifetime. Since these are heavier than the single beads,
their velocity after an impact is less affected by the liquid
bridges. Furthermore, the interior of a (larger) cluster may act
as a sink for the liquid due to capillarity, resulting in a
somewhat reduced wetness of the beads outside the cluster. Hence
the 'effective' restitution coefficient, averaged over all
entities in the system, increases again. As a consequence, the BNE
is now reduced, and the phases become better miscible, as we
observe.

This model is in accordance with the observation that the overall
appearance of the material is not expected to change noticeably in
this regime: as more liquid is added, the average cluster size is
just increased. The clusters take up the excess liquid, due to the
increased average coordination number of the bridge network within
a cluster. In other words, excess liquid is stored in liquid
bridges within clusters, which are not active in the dissipation
process.

We thus have elucidated the interplay between segregation (BNE)
and a phase transition to viscoplasticity, and have given a robust
criterion for the latter. The phase boundary can be understood
from simple energy considerations.

Inspiring discussions with H. Herrmann, S. Luding, and S. Linz are
gratefully acknowledged. The authors thank the German Science
Foundation for financial support within the Priority Program
`Wetting and Structure Formation at Interfaces'. MMK acknowledges
the Alexander von Humboldt Foundation for generous funding.

\bibliography{apssamp}

\end{document}